\documentclass[twocolumn,preprintnumbers,amsmath,amssymb]{revtex4-1}

\usepackage{graphicx}
\usepackage{dcolumn}
\usepackage{bm,epsfig}
\usepackage{epstopdf}

\begin{document}

\title{Unusual Kondo-hole effect and crystal-field frustration in Nd-doped CeRhIn$_{5}$}

\author{P. F. S. Rosa$^{1,2}$, A. Oostra$^{1}$, J. D. Thompson$^{2}$, P. G. Pagliuso$^{3}$, and Z. Fisk$^{1}$}

\affiliation{
$^{1}$ University of California, Irvine, California 92697-4574, U.S.A. \\
$^{2}$ Los Alamos National Laboratory, Los Alamos, New Mexico 87545, U.S.A.\\
$^{3}$ Instituto de F\'isica \lq\lq Gleb Wataghin\rq\rq, UNICAMP, Campinas-SP, 13083-859, Brazil.}
\date{\today}

\begin{abstract}
We investigate single crystalline samples of Ce$_{1-x}$Nd$_{x}$RhIn$_{5}$  by means of X-ray diffraction, microprobe, magnetic susceptibility, heat capacity, and 
electrical resistivity measurements. Our data reveal that the antiferromagnetic 
transition temperature of CeRhIn$_{5}$, $T_{N}^{\mathrm{Ce}} = 3.8$~K, is linearly suppressed with $x_{\mathrm{Nd}}$, by virtue of the
``Kondo hole" created by Nd substitution. The extrapolation of $T^{\mathrm{Ce}}_{N}$ to zero 
temperature, however, occurs at $x_{c} \sim 0.3$, which is below the 2D percolation limit found in Ce$_{1-x}$La$_{x}$RhIn$_{5}$. This result strongly suggests the presence of crystal-field frustration effects. Near$x_{\mathrm{Nd}} \sim 0.2$, the Ising AFM order from Nd ions is stabilized and $T^{\mathrm{Nd}}_{N}$ increases up to $11$~K in pure NdRhIn$_{5}$. Our results shed light on the  effects of magnetic doping in heavy-fermion antiferromagnets and stimulate the study of such systems under applied pressure.

 \end{abstract}

\maketitle

\section{INTRODUCTION}

A remarkable variety of unexpected phenomena arises when magnetic impurities are introduced in a metal. One of the most striking 
examples is that of Gold (Au) metal, which is, in its pure form, a good conductor displaying decreasing
electrical resistance with decreasing temperature. A few parts per million (ppm) of magnetic Iron impurities in Au, however, cause the resistance to rise 
logarithmically at a material-specific Kondo impurity temperature $T_{K}$ \cite{AuFe}. This so-called single-ion Kondo effect reflects the incoherent  
scattering of conduction electrons by the magnetic impurities introduced into the host \cite{Kondo}.  
In addition to introducing only a few ppm of magnetic impurities, 
it is also possible to synthesize materials in which there is a periodic lattice of Kondo ``impurities" coupled to the surrounding sea of itinerant electrons. At temperatures well above $T_{K}$, conduction electrons are scattered incoherently by the periodic array of Kondo ``impurities", but at much lower temperatures, translational invariance of the ``impurities" requires that a strongly renormalized Bloch state develop in which scattering is coherent \cite{ReviewKondo}.

A good example of these behaviors is found when La is replaced systematically by magnetic (Kondo) Ce ions in the 
non-magnetic host LaCoIn$_{5}$ \cite{Nakatsuji2002}. In the Kondo-lattice limit CeCoIn$_{5}$, the low temperature resistivity 
is very small, and the compound hosts unconventional superconductivity in which the effective mass of electrons in the 
renormalized conduction bands is large \cite{dhva}. If a small number of Ce atoms in CeCoIn$_{5}$ now is replaced by
 non-magnetic La ions, a Kondo-impurity effect develops on the La ions. The absence of a Ce ion in the periodic Kondo lattice 
 creates a ``Kondo-hole" such that La acts as a Kondo impurity and incoherently scatters electrons in the renormalized heavy conduction 
bands \cite{KHEric} In the series Ce$_{1-x}$La$_{x}$CoIn$_{5}$, the system evolves at low temperatures as a function of $x$ 
from a coherent Kondo lattice ($x=0$) to a a collection of incoherently scattering Kondo impurities at $x \sim 0.4$. 
Interestingly, this cross-over coincides with the percolation limit of a 2D square lattice on which the La/Ce ions sit.

Magnetic doping has not been extensively studied in CeCoIn$_{5}$ or other members within the Ce$_{m}M_{n}$In$_{3m+2n}$ ($M$ = transition metals Co, Rh, Ir, Pd, Pt; $n=0,1,2$; $m=1,2,3$) family of which it is a part. Recently, however, Nd-doping in CeCoIn$_{5}$ at concentrations $5-10$\% has been shown to induce an unexpected magnetic state inside the zero-field superconducting (SC)
phase \cite{CeNdPetrovic}. Remarkably, the  propagation vector and moment size of the incommensurate magnetic order in Ce$_{0.95}$Nd$_{0.05}$CoIn$_{5}$ are identical to those observed in the field-induced 
magnetically ordered phase in pure CeCoIn$_{5}$ ($Q$-phase) \cite{CeNdNeutrons,QPhase}. These results indicate that the Nd ions are fundamentally changing the electronic system and not acting simply as a Kondo hole. 
To our knowledge, there is no report on the effects of magnetic doping in the antiferromagnetic (AFM) member CeRhIn$_{5}$  up to date. In the following we address the open questions: (i)  how will Nd interact with the AFM order 
of CeRhIn$_{5}$? (ii) will there be a ``Kondo-hole" effect? We note that ``Kondo-hole" behavior has been observed in Ce$_{1-x}$La$_{x}$RhIn$_{5}$ where AFM order decreases linearly with La. In the limit $T_{N} \rightarrow 0$, the critical La concentration, $x_{c}\sim 40\%$, also reflects the percolation limit of the 2D square lattice \cite{CeLaPagliuso}.

In this work, we report the first study of magnetic doping in CeRhIn$_{5}$ by means of X-ray diffraction, microprobe, magnetization, heat capacity, and electrical resistivity measurements. Our data show that the AFM ordering temperature of CeRhIn$_{5}$  ($T_{N}^{\mathrm{Ce}} = 3.8$~K) decreases linearly with Nd concentration, $x_{\mathrm{Nd}}$, and extrapolates to zero at a critical Nd concentration of $x_{c} \sim 30\%$. Hence, in the dilute regime, Nd ions behave as a free paramagnetic impurity, i.e. a moment-bearing ``Kondo-hole" in the Ce system. The fact that  $x_{c}$ is below the percolation limit indicates that there is another mechanism frustrating the magnetic order
of the Ce sublattice. We argue that this mechanism is the crystal-field frustration due to the different spin configurations of 
CeRhIn$_{5}$ (easy $c$-axis magnetization but with ordered moments in-plane) and NdRhIn$_{5}$ (Ising spins along c-axis). In fact, around
 $x_{\mathrm{Nd}}
\sim 0.2$, the Ising AFM order of the Nd sublattice is stabilized and $T^{\mathrm{Nd}}_{N}$ increases up to $11$~K in pure NdRhIn$_{5}$.

\section{EXPERIMENTAL DETAILS}

Single crystalline samples of Ce$_{1-x}$Nd$_{x}$RhIn$_{5}$ ($x~=~0,0.05,0.1,0.15,0.2,0.3,0.5,0.7,0.9,1$) were grown by the In-flux technique. The crystallographic structure was verified by X-ray powder diffraction at room temperature. In addition, several samples were characterized by elemental analysis using a commercial Energy Dispersive Spectroscopy (EDS) microprobe. 

Magnetization measurements were performed using a commercial superconducting quantum interference device (SQUID). The specific heat was measured using a commercial small mass calorimeter that employs a quasi-adiabatic thermal relaxation technique. The in-plane electrical resistivity was obtained using a low-frequency ac resistance bridge and a four-contact configuration.

\section{RESULTS}

\begin{figure}[!ht]
\begin{center}
\hspace{-0.35cm}
\includegraphics[width=1.\columnwidth]{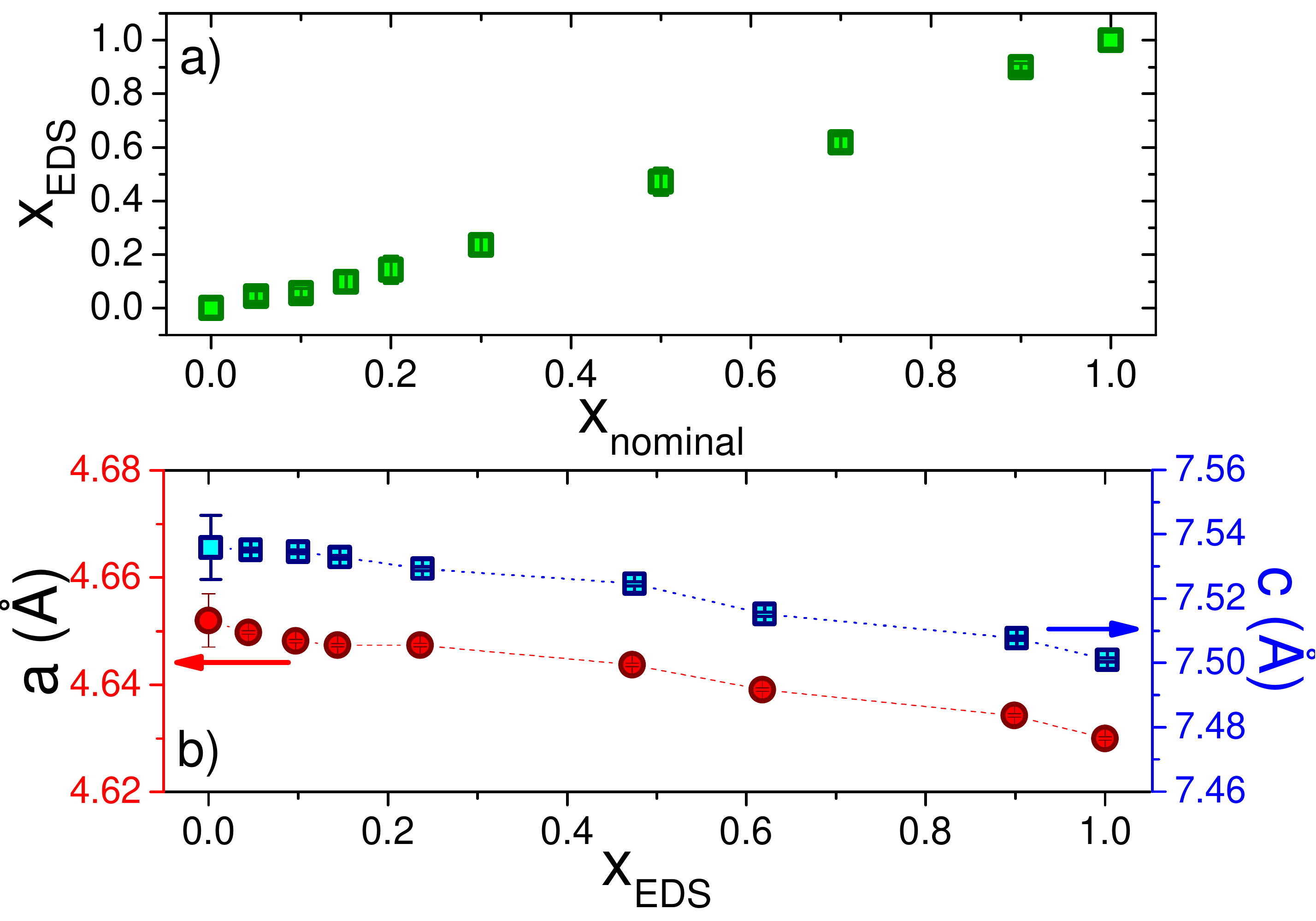}
\vspace{-0.5cm}
\end{center}
\caption{a) Actual concentration measured by EDS, $x_{EDS}$, as a function of nominal concentration, $x_{\mathrm{nominal}}$ 
in the series Ce$_{1-x}$Nd$_{x}$RhIn$_{5}$. b) Tetragonal lattice parameters as a function of $x_{EDS}$ along the series 
Ce$_{1-x}$Nd$_{x}$RhIn$_{5}$.} 
\label{fig:Fig1}
\end{figure}

Figure~\ref{fig:Fig1}a shows the actual Nd concentration obtained by EDS ($x_{\mathrm{EDS}}$) as a function of the nominal Nd concentration $x_{\mathrm{nominal}}$. The smooth and monotonic relationship between the $x_{\mathrm{EDS}}$ and $x_{\mathrm{nominal}}$ indicates that Nd is being incorporated in the lattice. Further, the small error bars, $\Delta x$, point to a rather homogeneous distribution of Nd. 
 In the extremes of the series, $x_{EDS}$ has an error bar of $\Delta x \sim 0.02$. For Nd concentrations around $50\%$, a larger variation of $\Delta x = 0.05$ is observed, which is expected for concentrations in the middle of the series. We note, however, that $\Delta x$ is the standard deviation accounting  for different samples 
from the same batch and not for a single sample. On average, the variation within a single crystal ($\sim 0.01$) is smaller than the standard deviation. These results indicate that Nd substitutes Ce homogeneously instead of producing an intergrown of NdRhIn$_{5}$. Herein, we will refer to the actual EDS concentration.

Figure~\ref{fig:Fig1}b shows the lattice parameters obtained by powder X-ray diffraction as a function of Nd concentration. The X-ray powder patterns show that all members of the series crystallize in the tetragonal 
HoCoGa$_{5}$ structure and no additional peaks are observed. A smooth decrease is found in both lattice parameters $a$ and $c$, in agreement with Vegard's law. This result implies that the volume of the unit cell is decreasing
with Nd concentration, suggesting that Nd doping produces positive chemical pressure. Using the bulk modulus of CeRhIn$_{5}$, we estimate that a rigid shift of the lattice parameters from CeRhIn$_{5}$ to 
Ce$_{0.95}$Nd$_{0.05}$RhIn$_{5}$ corresponds to $\Delta P = 0.25$~GPa of applied pressure. From the phase diagram of CeRhIn$_{5}$ under pressure \cite{TusonNature}, this $\Delta P$ would correspond to an increase of $T_{N}$ by  $0.1$~K. We will see below that the AFM order actually is suppressed in Ce$_{0.95}$Nd$_{0.05}$RhIn$_{5}$, indicating that chemical pressure is not the main tuning parameter determining $T_{N}$. 

\begin{figure}
\begin{center}
\hspace{-0.65cm}
\includegraphics[width=0.85\columnwidth,keepaspectratio]{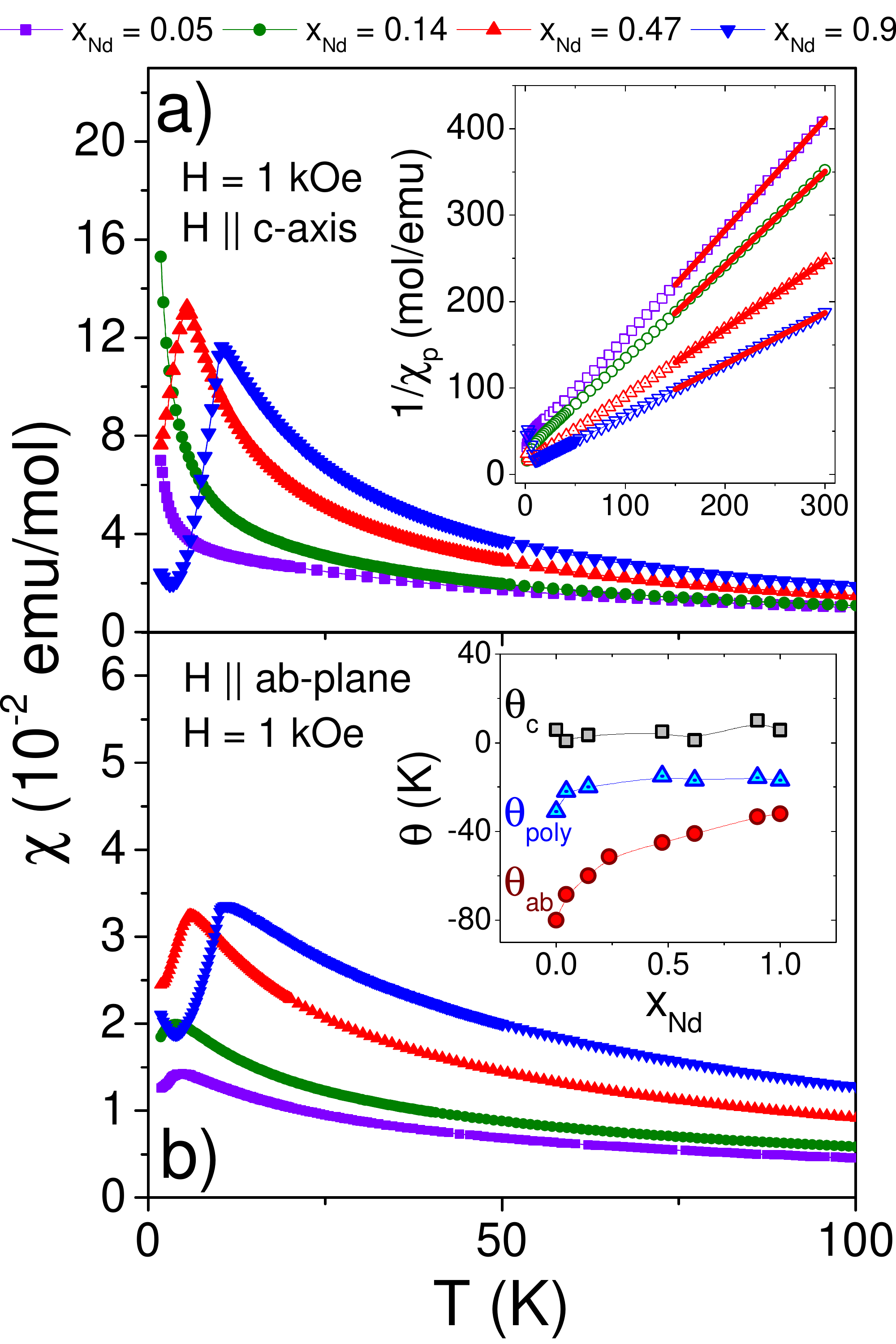}
\vspace{-0.5cm}
\end{center}
\caption{a) Temperature dependence of the magnetic susceptibility, $\chi_{c}(T)$, of representative samples in the Ce$_{1-x}$Nd$_{x}$RhIn$_{5}$ series in a field of $1$~kOe applied along the $c$-axis.
Inset shows the inverse susceptibility of the polycrystalline average $vs$ temperature. Solid lines are linear fits to the data.
b) Temperature dependence of the magnetic susceptibility, $\chi_{ab}(T)$, for the same samples in a field of $1$~kOe applied along the $ab$-plane.  
Inset shows the Curie-Weiss temperature, $\theta$, for all compositions of Ce$_{1-x}$Nd$_{x}$RhIn$_{5}$.}
\label{fig:Fig2}
\end{figure}

Figures~\ref{fig:Fig2}a and b show the $T$-dependence of the magnetic susceptibility, $\chi(T)$, for a field of $1$~kOe applied along the $c$-axis and $ab$-plane, respectively. For low Nd concentrations
($x_{\mathrm{Nd}}=0.05,\,0.14$), there is
no evidence of $T_{N}$ in the $\chi_{c}(T)$ data, i.e., when $H||c$-axis. This result is somewhat
unexpected because AFM order is observed by a clear peak in heat capacity measurements of both CeRhIn$_{5}$ and Ce$_{1-x}$Nd$_{x}$RhIn$_{5}$. 
Instead of an expected peak in $\chi(T)$, we observe a low-$T$ Curie-tail, suggesting that the Nd ions are free paramagnetic 
impurities embedded in the Kondo lattice.  When $H||ab$-plane, however, $\chi_{ab}(T)$ displays a very similar 
behavior when compared to pure CeRhIn$_{5}$: there is a maximum in $\chi(T)$ followed by a kink at $T_{N}^{\mathrm{Ce}}$. We attribute this difference to the fact that the spins in NdRhIn$_{5}$ point along the $c$-axis and the magnetic susceptibility 
along this direction is much larger than the in-plane susceptibility.
Thus, $\chi_{ab}(T)$ data reveals a linear decrease of $T_{N}^{\mathrm{Ce}}=3.8$~K with $x_{\mathrm{Nd}}$ up to  $x_{\mathrm{Nd}}=0.14$. Between $x_{\mathrm{Nd}}=0.14$ and $x_{\mathrm{Nd}}=0.23$, the transition temperature
starts to increase again, suggesting that the AFM order due to Nd ions starts to develop at $T_{N}^{\mathrm{Nd}}$.
Though not obvious in these data, $\chi_{ab}(T)$ reaches a maximum at $T^{\mathrm{max}}_{\chi} > T_N^{\mathrm{Ce}}$ in CeRhIn$_{5}$ and lightly Nd-doped samples.
The temperature $T^{\mathrm{max}}_{\chi}$ also decreases with $x_{\mathrm{Nd}}$, from $\sim 7.5$~K in pure CeRhIn$_{5}$ to $\sim3.2$~K at $x_{\mathrm{Nd}}=0.23$. Evidence for $T^{\mathrm{max}}_{\chi}$, however, is lost for  $x_{\mathrm{Nd}} > 0.23$ due to the dominant contribution from the Nd AFM order. We will return to this analysis when discussing the phase diagram of Fig. 5. Finally, for higher Nd concentrations, both $\chi_{c}(T)$ and $\chi_{ab}(T)$ show AFM behavior of a typical local moment system.

 From fits of the polycrystalline average of the data (inset of Fig.~\ref{fig:Fig2}a) to a Curie-Weiss law, we obtain effective magnetic moments of  2.5(1) $\mu_{B}$, 2.7(1) $\mu_{B}$, 3.2(1) $\mu_{B}$, and 3.7(1) $\mu_{B}$ for
 $x_{\mathrm{Nd}}=0.05, 0.14, 0.47, 0.9$, respectively. These calculated values are in good agreement with the theoretical values of  2.59 $\mu_{B}$, 2.69 $\mu_{B}$, 3.05 $\mu_{B}$, and 3.52 $\mu_{B}$ , respectively.
We also obtain the paramagnetic Curie-Weiss temperature, $\theta_{\mathrm{poly}}$, which averages out crystal electrical field (CEF) effects. 
The inset of Fig.~\ref{fig:Fig2}b shows $\theta_{\mathrm{poly}}$ as well as $\theta_{c}$ and $\theta_{ab}$. In a molecular field approximation, $\theta_{\mathrm{poly}}$  is proportional to the effective exchange interaction, $J$, between rare-earth ions. The fact that $\theta_{\mathrm{poly}}$ is negative is in agreement with the AFM correlations
found in the series.
A reduction of $\theta_{\mathrm{poly}}$ is observed going from CeRhIn$_{5}$ ($\theta_{\mathrm{poly}}=-31$~K) to 
NdRhIn$_{5}$  ($\theta_{\mathrm{poly}}=-17$~K), which suggests within a molecular field model that $J$ also decreases along the series. As a consequence, 
this reduction in $J$ would be expected to decrease the AFM ordering temperature. The experimental data, however,
shows the opposite behavior: $T_{N}$ in NdRhIn$_{5}$ ($T_{N}^{\mathrm{Nd}}=11$~K) is almost three times larger than in 
CeRhIn$_{5}$ ($T_{N}^{\mathrm{Ce}}=3.8$~K). Moreover, in a Kondo latice like CeRhIn$_{5}$, $\theta_{\mathrm{poly}}$  also includes the AFM Kondo exchange that tends to reduce $T_{N}$ relative to that expected solely from the indirect Ruderman-Kittel-Kasuya-Yosida (RKKY) interaction \cite{Doniach}.  Because there is no Kondo effect in NdRhIn$_{5}$, the variation in $\theta_{\mathrm{poly}}$  with $x_{\mathrm{Nd}}$ implies a suppression of the Kondo contribution and  increased dominance of the RKKY interaction. This is reflected in the ratio $T_{N}/\theta_{\mathrm{poly}}$, which is $0.12$ in CeRhIn$_{5}$ and $0.65$ in NdRhIn$_{5}$. As illustrated in the inset of  Fig.~\ref{fig:Fig2}b, $\theta_{\mathrm{poly}}$ reaches a plateau between $x_{\mathrm{Nd}}=0.23$ and $0.47$, suggesting that Kondo interactions are essentially quenched before $x_{\mathrm{Nd}}=0.47$. Consequently, one might expect $T_N$ to increase initially as Nd replaces Ce and then to remain approximately constant for $x_{\mathrm{Nd}}>0.47$. As we will come to, this is not the case and $T_{N}$ is a non-monotonic function of Nd content.
The above discussion indicates that there is another relevant mechanism determining the magnetic 
ordering in the series Ce$_{1-x}$Nd$_{x}$RhIn$_{5}$. From the nearly constant values of $\theta_{c}$ and pronounced change of $\theta_{ab}$, which anisotropy is a consequence of CEF effects, it is reasonable to expect that CEF effects play an important role. In fact, from the high-temperature expansion of $\chi(T)$ we can readily observe 
that the main tetragonal CEF parameter, $B_{2}^{0} \propto (\theta_{ab}-\theta_{c})$,  systematically decreases with Nd concentration.

\begin{figure}[!ht]
\begin{center}
\includegraphics[width=0.85\columnwidth,keepaspectratio]{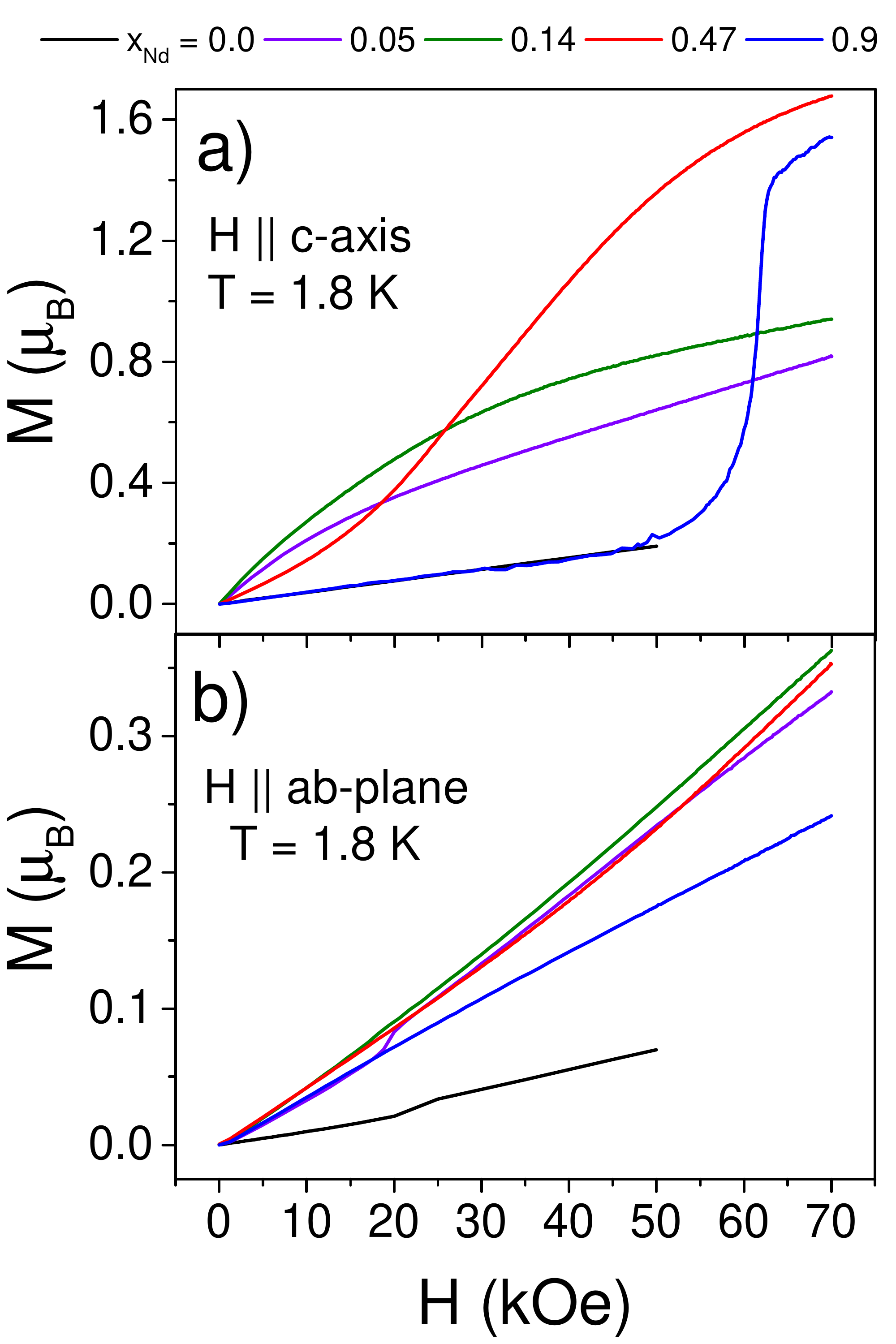}
\vspace{-0.7cm}
\end{center}
\caption{a) Field dependence of the magnetization at $1.8$~K for fields along the $c$-axis. Data for $x_{\mathrm{Nd}}=0$ coincide with that for $x_{\mathrm{Nd}}=0.9$ for $H \leq 50$~kOe.
b)  Field dependence of the magnetization at $1.8$~K for fields along the $ab$-plane.}
\label{fig:Fig3}
\end{figure}

Figures~\ref{fig:Fig3}a and b show the $H$-dependence of the magnetization, $M(H)$, at 1.8~K for fields applied along the $c$-axis and $ab$-plane, respectively. Although $M_{c}(H)$ for 
CeRhIn$_{5}$ displays a linear response with field,
 at low Nd concentrations ($x_{\mathrm{Nd}}=0.05,\,0.14$) there is a non-linear behavior that resembles
 a Brillouin function. This supports our interpretation of the origin of the low-$T$ Curie tail in $\chi_{c}$ for low Nd content, namely that Nd ions at low concentrations act as free paramagnetic 
 entities. Because the Brillouin-like contribution to $M_{c}(H)$ is substantially larger than expected from just a simple free Nd moment, this behavior implies that Nd moments also are locally inducing free-moment like character on neighboring Ce atoms. This is most pronounced for $H||c$ due to the much higher susceptibility of Nd moments along this direction. At light Nd doping, then, Nd acts as a rather different type of ``Kondo hole" compared to that induced by non-magnetic La substitution for Ce. The Nd ions carry a net magnetic moment that is not quenched by the Kondo-impurity effect. At high $x_{\mathrm{Nd}}$,  $M_{c}(H)$ displays a field-induced transition to a spin-polarized state, as observed in NdRhIn$_{5}$.
 When the field is along the ab-plane, pure CeRhIn$_{5}$ also displays a weak field-induced anomaly in $M_{ab}(H)$ ($H_{c} \sim 22$~kOe), which signals a change in ordering wavevector \cite{Flouquet} and is suppressed with $x_{\mathrm{Nd}}$.

\begin{figure}[!ht]
\begin{center}
\hspace{-0.75cm}
\includegraphics[width=0.85\columnwidth,keepaspectratio]{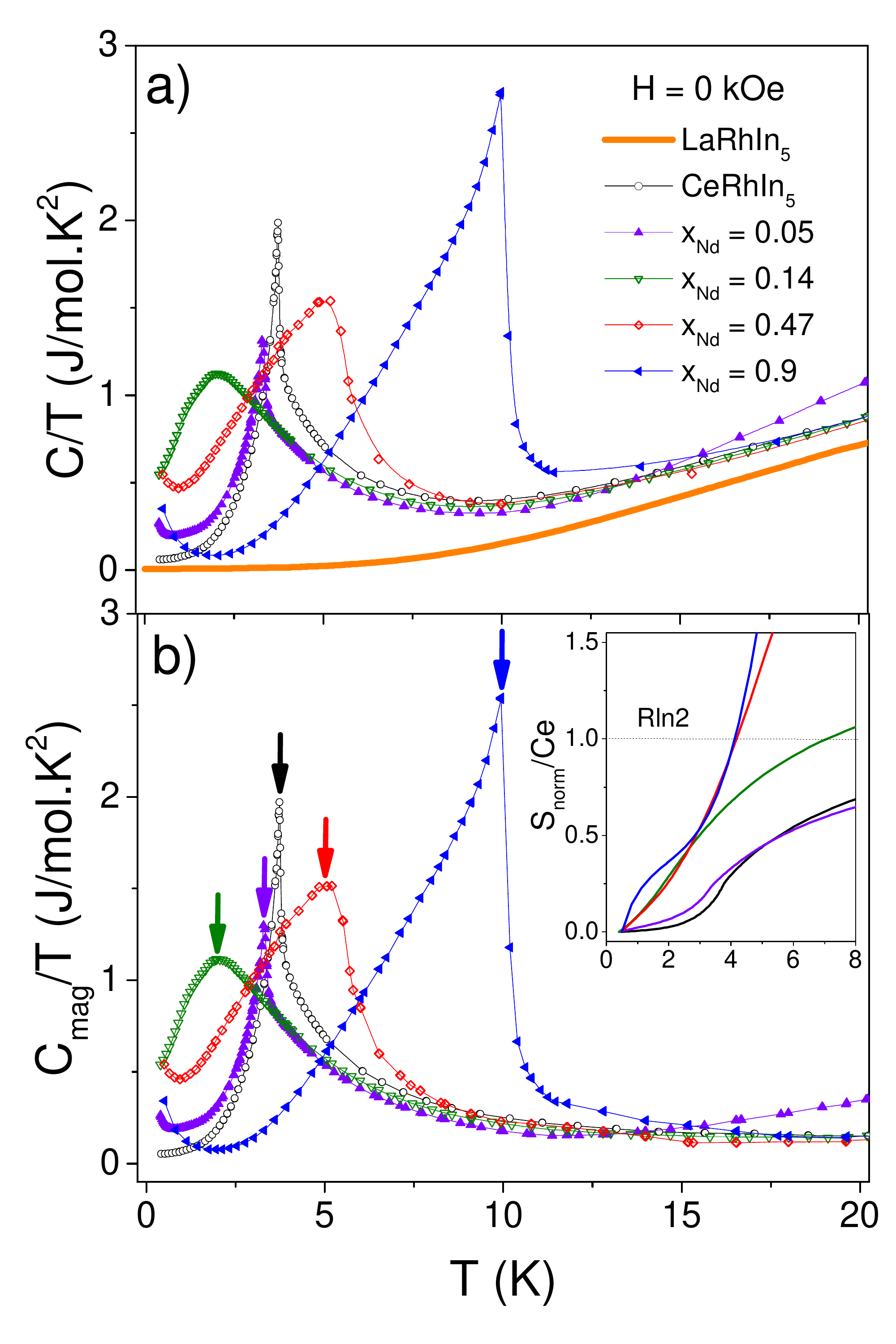}
\end{center}
\vspace{-0.7cm}
\caption{a) Temperature dependence of the specific heat, $C/T$, of LaRhIn$_{5}$, CeRhIn$_{5}$ and representative samples of
the series Ce$_{1-x}$Nd$_{x}$RhIn$_{5}$. b) Magnetic contribution to the specific heat, $C_{\mathrm{mag}}/T$, as a function of temperature. Inset shows the entropy per Ce normalized by $R$ln$2$.}
\label{fig:Fig4}
\end{figure}

Figure~\ref{fig:Fig4}a shows the temperature dependence of the heat capacity over temperature, $C/T$, for
four representative Nd concentrations. LaRhIn$_{5}$, the non-magnetic member, and pure CeRhIn$_{5}$
also are included. The sharp peak at  $T_{N}=3.8$~K displayed by CeRhIn$_{5}$ first decreases 
linearly with Nd concentrations up to $x_{\mathrm{Nd}}=0.14$. At $x_{\mathrm{Nd}}=0.14$, the transition at $T_{N}$ starts to broaden
and further increase in $x_{\mathrm{Nd}}$ reveals an enhancement of $T_{N}$, in agreement with $\chi(T)$ data.

Figure~\ref{fig:Fig4}b shows the magnetic contribution to the heat capacity, $C_{\mathrm{mag}}/T$, after subtracting
 LaRhIn$_{5}$ from the data. The transition temperature at which $C_{\mathrm{mag}}/T$ peaks is marked by the arrows. As the 
temperature is lowered further, an upturn is observed for all crystals with finite $x_{\mathrm{Nd}}$, including NdRhIn$_{5}$, suggesting that the Nd 
ions are responsible for it. Reasonably, the upturn may be associated with the nuclear moment of Nd ions, and it can be fit well by a sum of both electronic ($\propto \gamma$) and nuclear ($\propto T^{-3}$) terms \cite{SchottkyNuclear}, consistent with the presence of a nuclear Schottky contribution. 

The magnetic entropy as a function of 
temperature is obtained by integrating $C_{\mathrm{mag}}/T$ over $T$. The inset of Figure~\ref{fig:Fig3}b shows the $T$-
dependence of the magnetic entropy recovered per Ce ion. The entropy is normalized by 
$R$ln2, which is the entropy of the ground state doublet. In pure CeRhIn$_{5}$ (black bottom curve), the 
magnetic entropy increases with $T$ followed by a kink at $T_{N}$.
 We observe an increase in the recovered entropy below $T_{N}$ even when a very small amount of Nd is introduced (e.g., $x_{\mathrm{Nd}}=0.05$). Increasing the concentration to $x_{\mathrm{Nd}}=0.14$ yields a further entropy increase. This result indicates that the magnetic entropy does not scale with the Ce concentration, in turn suggesting  that the extra magnetic entropy comes from the free paramagnetic Nd ions. 
 
 \begin{figure}[!ht]
\begin{center}
\hspace{-0.7cm}
\includegraphics[width=1.05\columnwidth,keepaspectratio]{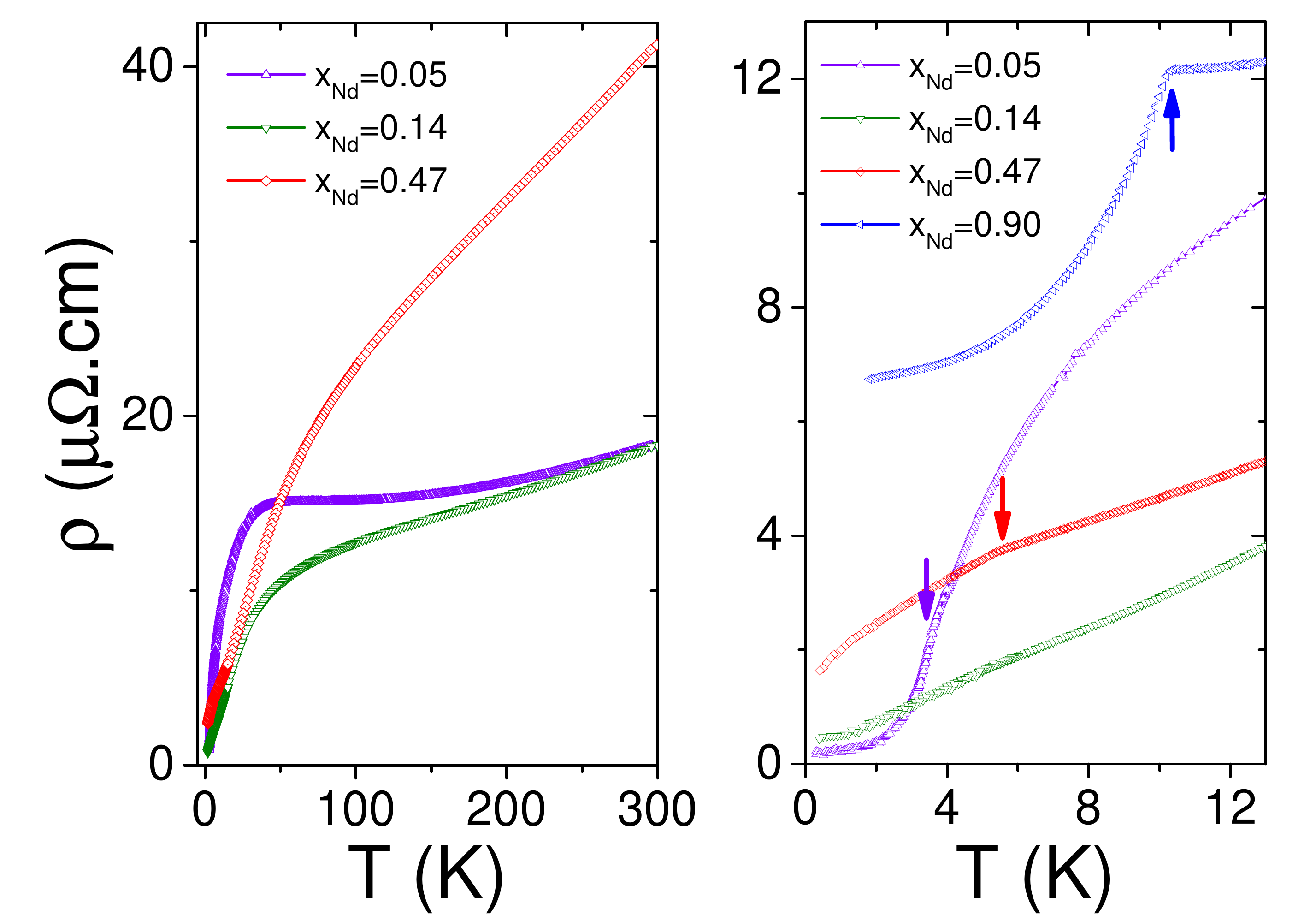}
\end{center}
\caption{a) In-plane electrical resistivity, $\rho_{ab}(T)$, of Ce$_{1-x}$Nd$_{x}$RhIn$_{5}$ as a function of temperature. b) Low temperature  $\rho_{ab}(T)$ data. Arrows mark $T_{N}$.}
\label{fig:Fig6}
\end{figure}

Finally, we discuss the temperature dependence of the in-plane electrical resistivity, $\rho_{ab}(T)$, of 
Ce$_{1-x}$Nd$_{x}$RhIn$_{5}$. Figure~5a shows $\rho_{ab}(T)$ for samples with $x_{\mathrm{Nd}}<0.5$. At
 $x_{\mathrm{Nd}}=0.05$, $\rho_{ab}(T)$ is very similar in magnitude and $T$-dependence to that of pure CeRhIn$_{5}$.
In particular, the broad peak at $\sim 40$~K indicates the crossover from incoherent Kondo scattering
at high temperatures to the heavy-electron state at low temperatures. As $x_{\mathrm{Nd}}$ is increased, $\rho_{ab}(T)$ decreases monotonically with temperature and the initial peak turns into a broad feature around $70$~K when $x_{\mathrm{Nd}}=0.47$. We 
note that the second CEF excited state of NdRhIn$_{5}$ is near 68~K, suggesting that the broad feature in $\rho_{ab}(T)$ is likely associated with CEF depopulation \cite{OnukiRRhIn5}. This evolution is consistent with an increase in the local character of the $4f$ system. Further, the low-temperature data shown in Fig.~5b display an increase in $\rho_{0}$. Typically disorder scattering would be expected to be a maximum near $x=0.5$, but this is not the case. As shown in Ref.~\cite{OnukiRRhIn5}, the residual resistivity of pure NdRhIn$_{5}$ is much lower than that of our Ce$_{0.1}$Nd$_{0.9}$RhIn$_{5}$ crystal. This difference implies that spin-disorder scattering plays a significant role in determining $\rho_{0}$ in this series.

\section{DISCUSSION}

 In Figure~\ref{fig:Fig6} we summarize our results in a $T-x$ phase diagram, in which two distinct regimes become clear. 
   The first one, at low Nd concentrations, presents a linear decrease of $T_{N}$ with $x_{\mathrm{Nd}}$. Interestingly, a 
 linear dependence of $T_{N}$ also has been observed in Ce$_{1-x}$La$_{x}$RhIn$_{5}$, where La creates a 
 ``Kondo hole'' in the system via dilution. In the La-doping case, however, $T_{N}$ extrapolates to $T=0$ at 
 a critical concentration of $x_{c}\sim 40\%$, which is the percolation limit of the $2D$ lattice. Here, Nd-doped 
 CeRhIn$_{5}$ displays a smaller $x_{c}$ of $\sim 30\%$, indicating that there is an additional  mechanism that frustrates N\'eel order
in the Ce sublattice \cite{CeLaPagliuso}.

\begin{figure}[!ht]
\begin{center}
\hspace{-0.75cm}
\includegraphics[width=1\columnwidth,keepaspectratio]{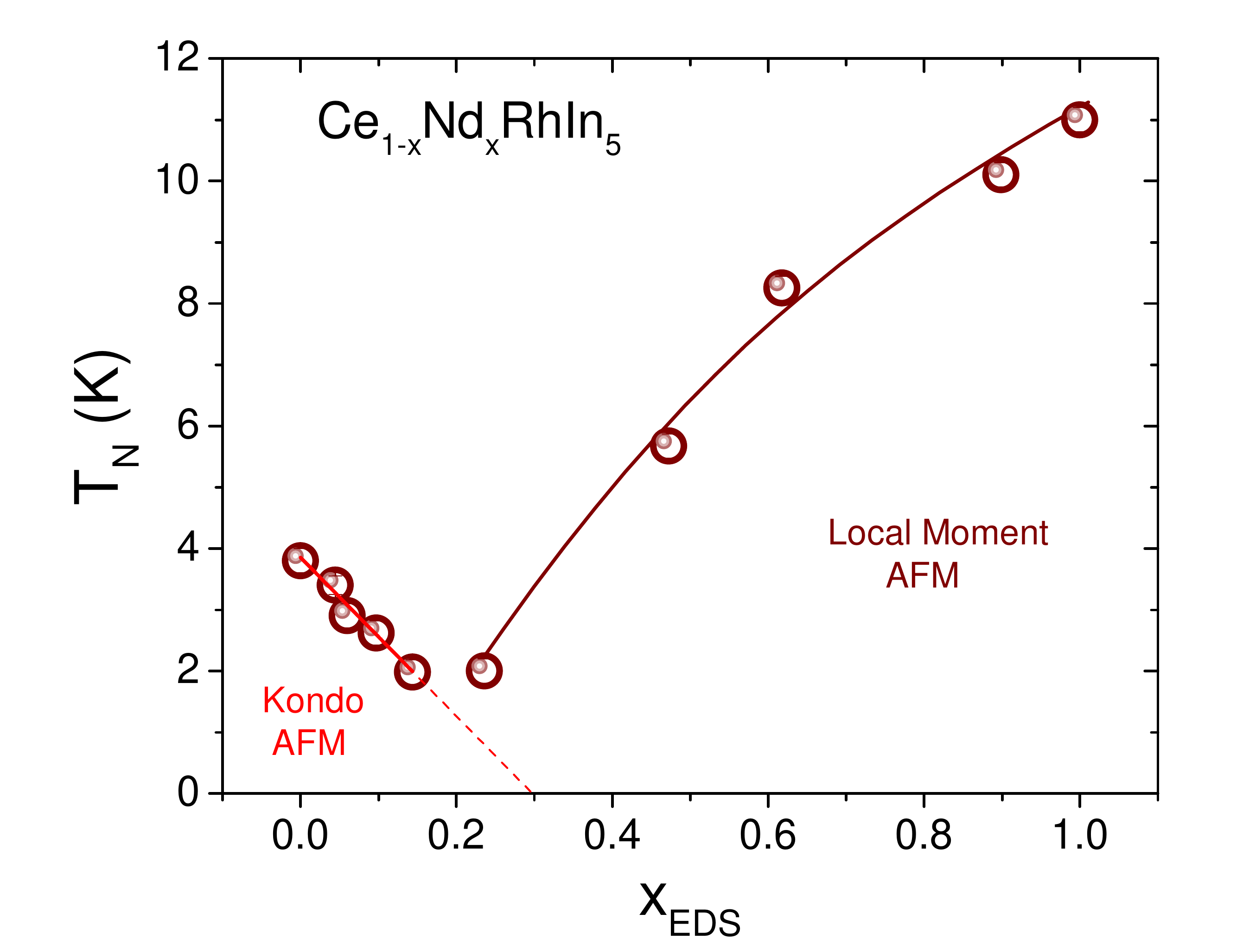}
\end{center}
\caption{$T-x$ phase diagram of the series Ce$_{1-x}$Nd$_{x}$RhIn$_{5}$.}
\label{fig:Fig6}
\end{figure}

It has been shown $-$ both theoretically and experimentally $-$ that $T_{N}$ in tetragonal structures is enhanced with respect to their cubic $R$In$_{3}$ ($R=$ rare-earth ions) counterparts whenever $R$ has an Ising magnetic structure, i.e., spins polarized along the $c$-axis \cite{PG2006,SerranoTb}. This is due to the fact that the tetragonal CEF parameters in these structures favor a groundstate with Ising symmetry, as supported by the fact that the $c$-axis susceptibility is  larger than the $ab$-plane susceptibitility in members whose $R$ element has finite orbital momentum. Because NdRhIn$_{5}$ displays commensurate Ising-like order below $T_{N}=11$~K, it is reasonable
to assume that Nd$^{3+}$ ions will retain their Ising-like character when doped into the Ce sites \cite{NdRhIn5MagStruc}. CeRhIn$_{5}$, however, has an incommensurate magnetic structure with spins perpendicular to the $c$-axis \cite{CeRhIn5MagStruc}. Hence, a crystal-field
frustration of the in-plane order in CeRhIn$_{5}$ is induced by Nd$^{3+}$ Ising spins. As a consequence, $T_{N}^{\mathrm{Ce}}$ of the
Ce sublattice extrapolates to zero before the percolation limit and $T_{N}^{\mathrm{Nd}}$ of the Nd sublattice is stabilized.

\vspace{1cm}
\section{CONCLUSIONS}

In summary, we synthesize single crystals of Ce$_{1-x}$Nd$_{x}$RhIn$_{5}$ 
using the In-flux technique. X-ray diffraction and microprobe measurements show
a smooth evolution of lattice parameters and Nd concentration, respectively. Across the doping series, there is a complex interplay among Kondo-like impurity physics, 
magnetic exchange and crystal-field effects as the Nd content changes.  At low $x_{\mathrm{Nd}}$, there is an unusual type of magnetic ``Kondo hole"
and $T^{\mathrm{Ce}}_{N}$ decreases linearly with $x_{\mathrm{Nd}}$. The extrapolation of $T^{\mathrm{Ce}}_{N}$ to zero 
temperature occurs below the 2D percolation limit due to crystal-field frustration effects. Around
 $x_{\mathrm{Nd}}
\sim 0.2$, the Ising AFM order from Nd ions is stabilized and $T^{\mathrm{Nd}}_{N}$ increases up to $11$~K in pure NdRhIn$_{5}$.
 Further investigation of the Ce$_{1-x}$Nd$_{x}$RhIn$_{5}$ series under pressure will be valuable to understand this interplay in the superconducting state.

\begin{acknowledgments}
Work at Los Alamos was performed under the auspices of the U.S. Department of Energy, Office of Basic Energy Sciences, Division of Materials Science and Engineering. P. F. S. R. acknowledges: (i) a Director's Postdoctoral Fellowship though the LANL LDRD program; (ii) FAPESP Grant 2013/20181-0.

\end{acknowledgments}

\bibliography{basename of .bib file}

\begin{thebibliography}{99}

 \bibitem{AuFe} W. J. De Haas, G. J. Van Den Berg, Physica \textbf{3}, 6 440-449 (1936).
 \bibitem{Kondo} J. Kondo, Prog. Theor. Phys. 32:37-49 (1964).
 \bibitem{ReviewKondo} G. R. Stewart, Rev. Mod. Phys. 56, 755 (1984).
 \bibitem{Nakatsuji2002} S. Nakatsuji, S. Yeo, L. Balicas, Z. Fisk, P. Schlottmann, P. G. Pagliuso, N. O. Moreno, J. L. Sarrao, and J. D. Thompson, Phys. Rev. Lett. \textbf{89}, 106402 (2002).
 \bibitem{dhva} R. Settai \textit{et al}, J. Phys. Condens. Matter \textbf{13}, L627-L634 (2001).
 \bibitem{KHEric} E. D. Bauer \textit{et al.}, Proc. Natl. Acad. Sci. USA \textbf{108} 6857 (2011).
 \bibitem{CeNdPetrovic} R. Hu, Y. Lee, J. Hudis, V. F. Mitrovic, and C. Petrovic, Phys. Rev. B \textbf{77}, 165129 (2008).
\bibitem{CeNdNeutrons} S. Raymond, S. M. Ramos, D. Aoki, G. Knebel, V. P. Mineev, and G. Lapertot, J. Phys. Soc. Jpn. \textbf{43}, 013707 (2014).
\bibitem{QPhase} M. Kenzelmann, S. Gerber, N. Egetenmeyer, J. L. Gavilano, Th. Strassle, A. D. Bianchi, E. Ressouche, R. Movshovich, E. D. Bauer, J. L. Sarrao, and J. D. Thompson,
Phys. Rev. Lett. \textbf{104}, 127001 (2010).
\bibitem{CeLaPagliuso} P. G. Pagliuso, N. O. Moreno, N. J. Curro, J. D. Thompson, M. F. Hundley, J. L. Sarrao, Z. Fisk, A. D. Christianson, A. H. Lacerda, B. E. Light, and A. L. Cornelius, Phys. Rev. B \textbf{66}, 054433 (2002).
\bibitem{TusonNature} T. Park, F. Ronning, H. Q. Yuan, M. B. Salamon, R. Movshovich, J. L. Sarrao and J. D. Thompson. Nature \textbf{440}, 65-68 (2006).
\bibitem{Doniach} S. Doniach, Physica B \textbf{91}, 231 (1977).
\bibitem{Flouquet} S. Raymond, E. Ressouche, G. Knebel, D. Aoki and J Flouquet, J. Phys.: Condens. Matter \textbf{19}, 242204 (2007).
\bibitem{SchottkyNuclear} A. C. Anderson, B. Holmstrom, M. Krusius, and G. R. Pickett, Phys. Rev. \textbf{183}, 546 (1969).
\bibitem{OnukiRRhIn5} N. V. Hieu \textit{et al}, J. Phys. Soc. Jpn. \textbf{76}, 064702 (2007).
\bibitem{PG2006} P. G. Pagliuso, D. J. Garcia, E. Miranda, E. Granado, R. Lora Serrano,
C. Giles, J. G. S. Duque, R. R. Urbano, C. Rettori, J. D. Thompson, M. F. Hundley and J. L. Sarrao,
J. Appl. Phys. \textbf{99}, 08P703 (2006).
\bibitem{SerranoTb} R. Lora-Serrano, C. Giles, E. Granado, D. J. Garcia, E. Miranda, O. Ag\"uero, L. Mendon\c{c}a-Ferreira, J. G. S. Duque, and P. G. Pagliuso, Phys. Rev. B \textbf{74}, 214404 (2006).
\bibitem{NdRhIn5MagStruc} S. Chang, P. G. Pagliuso, W. Bao, J. S. Gardner, I. P. Swainson, J. L. Sarrao, and H. Nakotte, Phys. Rev. B \textbf{66}, 132417 (2002).
\bibitem{CeRhIn5MagStruc} W. Bao, P. G. Pagliuso, J. L. Sarrao, J. D. Thompson, Z. Fisk, J. W. Lynn, and R. W. Erwin, Phys. Rev. B \textbf{62}, R14621(R) (2000). 




\end{thebibliography}

\end{document}